# The state-of-the-art in web-scale semantic information processing for cloud computing

**Wei Yu**
School of Information Management, Nanjing University, P. R. China
yuw.nju@gmail.com
**Junpeng Chen**
School of Computer Science, Wuhan University, P. R. China

**Abstract**   Based on integrated infrastructure of resource sharing and computing in distributed environment, cloud computing involves the provision of dynamically scalable and provides virtualized resources as services over the Internet. These applications also bring a large scale heterogeneous and distributed information which pose a great challenge in terms of the semantic ambiguity. It is critical for application services in cloud computing environment to provide users intelligent service and precise information. Semantic information processing can help users deal with semantic ambiguity and information overload efficiently through appropriate semantic models and semantic information processing technology. The semantic information processing have been successfully employed in many fields such as the knowledge representation, natural language understanding, intelligent web search, etc. The purpose of this report is to give an overview of existing technologies for semantic information processing in cloud computing environment, to propose a research direction for addressing distributed semantic reasoning and parallel semantic computing by exploiting semantic information newly available in cloud computing environment.

**Keywords**   Cloud computing. Web-scale semantic information processing. Distributed semantic reasoning. Parallel semantic computing

## 1 Introduction

With the increasing popularity of Internet services such as web search, instant message, e-commerce platforms and image processing, and various other types of third-party services on the Web, millions users give millions of clicks to acquire information every day. This brings terabytes of valuable data to be used to improve online performance and the demand for real-time applications and high-speed data processing. It also causes a critical bottleneck for many enterprises to construct large datacenter to satisfy the information need.

As a solution to these problems, cloud computing technology emerged and grows quickly. The companies who consume IT services or provide web services no longer need large capital outlays in hardware and software to deploy their services. They can just purchase these "hardware" and "software" as cloud computing services on the Web. These cloud computing services are quite cheap compared to the expense to build these infrastructures by one's own. Besides this economic advantage, cloud computing can provide on-demand self-service, broad network access, resource pooling, rapid elasticity and measured service [1]. This paradigm shift is transforming the IT industry. Developers with innovative ideas on web service are more easily to implement their applications and the web services are more efficient so as to attract more users.

These cloud computing applications and human users generate rich information and various kinds of knowledge which have never been processed before. Since most of cloud computing services are building on distributed web environment and provides virtualized resources, they bring large scale heterogeneous and distributed information which pose a great challenge in terms of the semantic ambiguity and also create many new research issues such as distributed metadata storage, distributed semantic reasoning, distributed semantic models construction, parallel semantic computing, etc. It also introduces many real world applications. For examples, cloud resource discovery, semantic analysis in science computing, web context understanding, and accurate advertise recommendation in commercial services. These research issues have been receiving growing attentions in web service field, data mining and among others in the recent years. Many significant researches have been done on these research topics. The main purpose of this document is to provide a survey of the development on these research challenges.

The rest of the report is organized as follows. The Sect. 2 begins with a brief introduction for cloud computing to provide the foundation of this concept. The background of cloud computing is given first, and then review some well-known cloud computing services. In Sect. 3, we review the origins of web-scale semantic information processing and explore the use of the semantic web technologies for improving the semantic information processing in the web environment. Sect. 4 presents overview of distributed semantic representation and some parallel semantic computing and applications. Section 5 concludes this study.

**2 Cloud computing**

Although the term cloud computing is being used in new ways since the availability of the dynamic and on-demand cloud services the technologies behind it are not new. The distributed architecture and virtualization technologies make sharing infrastructures, communication, data with others, and providing efficient and stable Internet services to users so that they can cut costs by eliminating the need for physical hardware and maintaining the software.

2.1 Definition of cloud computing

According to "The NIST Definition of Cloud Computing" [2], cloud computing is "the delivery of computing as a service rather than a product, whereby shared resources, software, and information are provided to computers and other devices as a utility (like the electricity grid) over a network (typically the Internet)". In these report, cloud computing is defined as "a style of computing over the Internet where dynamically scalable web services and virtualized resources are provided". In cloud computing everything can be provided as service, including infrastructure, hardware platform, software, etc. So the terms of the cloud computing and cloud service are used interchangeably in this report.

The cloud service model can be mainly classified into three layers [2]: Cloud Software as a Service (SaaS), Cloud Platform as a Service (PaaS), and Cloud Infrastructure as a Service (IaaS). In SaaS, the capability provided to the consumer is to use the provider's applications running on a Cloud Infrastructure. This is an alternative to locally run applications. The online word processor such as Google Docs is a typical SaaS. While in PaaS, users can have more choices and flexibility. The capability provided to the consumer is to deploy onto the cloud Infrastructure consumer-created or acquired application created using programming languages and tools supported by the provider. There are well-known PaaS examples such as the Google Apps Engine and Microsoft Azure. And in IaaS, the most flexibility and choices are provided. The consumer can chose processing, storage, networks, and other fundamental computing resources where the consumer is able to deploy and run arbitrary software, which can include operating systems and applications. Amazon's Elastic Compute Cloud (EC2) based on Amazon machine Image (AMI) is a kind of IaaS. These resources in IaaS are virtualized and dynamically scalable. The technology details of how to deploy and maintain these virtual resources are made in a transparent manner.

The operation of extremely large scale computer datacenters was the key enabler of cloud computing, as these datacenters take advantage of economies of scale, leading to decreases in the cost of electricity, bandwidth, operations, and hardware [1]. If the cloud service is based on Internet and available to the general public, it is referred as public cloud. When the cloud service is based on internal datacenters of a business or other organization and not made available to the general public, it is called private cloud. A hybrid cloud is a composition of these deployment models of the cloud computing. The cloud service providers can decide adopting which deployment models based on the application goals of the cloud service consummers. Given the advantages cloud computing has, it becomes the new technology trends and business model while more companies join to provide or consume cloud service.

2.2 Brief history of cloud computing

The term "cloud" is used as a metaphor for the Internet, based on the cloud drawing used to represent the computer network. The underlying concept of cloud computing dates back to the 1960s, when John McCarthy opined that "computation may someday be organized as a public utility." At the time Google started in 1998, its business increased so rapidly that the traditional IT technologies are not enough to process the huge amount of data in acceptable manner. To solve this problem, it implements its own file system Google File System (GFS) [3], and built its parallel computing environment MapReduce [4] and data storage Bigtable [5] based on GFS. These technologies which are later called "cloud computing" turned out to be high efficient as well as stable and reliable.

With the ubiquitous availability of high capacity networks, low cost computers and storage devices as well as the widespread adoption of virtualization, service-oriented architecture, autonomic, and utility computing have led to a tremendous growth in cloud computing. Another important cloud service provider, Amazon had found that their modern data centers were using as little as 10% of their capacity at any one time. Having found that the new cloud architecture resulted in significant internal efficiency improvements, Amazon initiated a new product development effort to provide cloud computing to external customers, and launched Amazon Web Service (AWS) on a utility computing basis in 2006. Elastic Computer Cloud (EC2) is constructed to provide scalable computing service on demand, and can be paid by hours. Amazon Simple Storage Service (S3) which store data in cloud infrastructure base on Dynamo [6] can be also bought by the using time. Besides Amazon, other companies also develop their own cloud computing services. IBM released "Blue Cloud" service while Microsoft calls its cloud service platform "Azure". eBay provides their own opensource PaaS platform turmeric. And Yahoo also develop data storage and processing platform "Sherpa" and non-structure data storage base "Mobstor".

In 2004, the most famous opersource cloud computing framework Hadoop began to build. It supports data-intensive distributed applications to work with thousands of computational independent computers and petabytes of data. Hadoop was derived from Google's MapReduce and Google File System (GFS) and is adopted by IBM, Yahoo and Facebook to construct cloud service infrastructures. In early 2008, Eucalyptus [7] became AWS API-compatible platform for deploying private clouds. At the same time, OpenNebula, an open-source cloud computing toolkit is enhanced in the RESERVOIR European Commission-funded project. Nowadays, more related projects are developed to support Hadoop framework and various technologies are applied. Pregel [8] is a new proposed cloud computing model for large scale graph processing. It can avoid multiple iteration in MapReduce framework and give stable and scalable performance. Pig-latin [9], a SQL-like and data flow language can be implemented on Pig [10] to perform database-like functionality.

2.3 In summary

Cloud computing is a type of virtual Internet service that has grown tremendously in popularity both in industry and academic over the past few years. Through cloud computing, users can process and store a large scale of data with only simple end while connect to Internet.

When users consume cloud service, they begin by logining to the websites who provide cloud services, then choose some of the cloud services they like. Users can choose different sizes of the virtual resources such as CPU processors, memories and hard disks to build a "bare" machine in the cloud to set up their applications. Or they can run their web services application on the platforms the cloud service producer provides. If users are tired of any implementation, they can just adopt the existing software services in the cloud. These cloud service can provide and consume lots of information on the web which may lead to ambiguation.

Given the success of semantic information processing in web service, such as semantic web search and knowledge representation, it is considered worthwhile to revisit semantic information processing problem through the novel perspective of cloud computing. In general, semantic information processing aim to help machine to understand and reasoning the semantics contained in text, web services, and other structured or unstructured data through shared semantic model and using semantic rules. However, no emphasis has been placed yet on semantic service based explicitly on cloud computing.

**3 Web-Scale semantic information processing**

Semantic information processing has been long studied as an essential problem in artificial intelligence (AI). The underlying of AI is to make machine "understand" human beings, and this is also a basic goal of the computer invitation. Semantic information processing can make "meaning" and the relations more obviously and formally for information sharing and knowledge discovery. Semantic information are used to classified into two kinds: pure semantic information that deals with the properties of artificially constructed formal system; descriptive semantic information, a empirical search for rules governing truth and meaning fullness of sentences in natural language [11]. On the web, the semantic information contains both of these two kinds. To understand and represent the pure semantic information, appropriate model should be build and formal standards are adopted. While for the second kind, metadata and semantic role can be used to describe and tag the meaning containing in natural language. In current semantic information processing, semantic models and natural language understanding technologies are both applied to make the web-scale semantic information obviously and sharing in different users and applications.

Great amount of web pages and formal system services are now accessible in cloud computing. Compared to the traditonal semantic information, the web-scale semantic information are often in large size and interrelated. To tacke the web-scale semantic information, metadata and tags are added to help information search and classify in order to provide users interesting information. In addition, semantic model are constructed to assist understanding text meaning. The richness of the web-scale semantic information challenges the current techniques and also provides new possibilities for accurately semantic information processing. Thus how to incorporate the new features and practices of cloud computing into semantic service becomes an important and urgent research topic.

In this section, we review the semantic web technologies first and then articulate some problems of current web-scale semantic information processing, and also indentify some new challenges to semantic service in cloud computing.

3.1 Semantic web

In order to process the large scale of semantic information on the web and to tackle with the semantic ambiguity, Tim Berners-Lee proposed semantic web as "a web of data that can be processed directly and indirectly by machines"[12]. The semantic web outperforms the traditional web for it can be interpreted by machines so that machines can "understand" and respond to complex human requests based on their meaning. Hence, machines can perform more of the tedious work involved in finding, combining, and acting upon information on the Web. The semantic web technologies have been widely applied in industry, biology and human science research [13].

In the semantic web, everything including resources, services, semantic relations, etc, can be identified with Uniform Resource Identifier (URI). This can avoid semantic ambiguities and is convenient to make version control. An elemental syntax is provided by XML to build content structure within document. The data interchange is mainly through Resource Description Framework (RDF) and Resource Description Framework Schema (RDFS). The former expresses data models referred to resources and their relationships. While the later is a vocabulary extending RDF to describe properties and classes of RDF-based resources. The Web Ontology Language (OWL) gives more vocabulary such as equality, richer typing of properties, characteristics of properties, etc to describe properties and classes. A RDF query language, SPARQL, can provide query services for semantic web data sources. These semantic web technologies can make knowledge unambiguous and serve to both intelligent agents and human beings. For example, the semantic wikis[14], which adopt the knowledge model and get the formal reasoning ability from the semantic web, can capture or identify information about the data within and between pages, thus provide more effective knowledge discovery and search service than traditional wikis.

By adopting a series of standards for metadata and vocabularies to describe resources and semantic relations, the semantic web can explain the structure of the knowledge and enable semantics to be added to the content in a machine-readable way. For example, the friend of a friend (FOAF), a popular vocabulary on the semantic web, uses RDF to describe the relationships people have to other people and the "things" around them. FOAF permits intelligent agents to choose the connections among thousands of relations between people enumerated by traditional web search engines. Also, the vast number of connections may be hard for human interpretation to analyze.

The goal of the semantic web is the Web of data. The linked data [15], as a good way to practice the semantic web, can help users and applications reach data and resources on the web directly through URL. Hence, in the cloud computing environment, the semantic web applications are more likely to be kinds of SaaS or PaaS. The semantic web technologies can help organize metadata and represent domain knowledge in the cloud computing. It can also solve the semantic ambiguity and heterogeneity problems brought by the distributed architecture and big data in cloud computing services and hence encourage data sharing and knowledge discovery. One of the linked data resources, DBpedia [16], constructs its service based on the cloud computing platform provided by Amazon. DBpedia takes all advantages of the efficient and stable cloud computing platform and the unambiguous and easy-accessible linked open data to provide semantic search and knowledge discovery in cloud computing environment. Compared to DBpedia, OCLC (Online Computer Library Center) developed its own cloud computing service Duracloud platform. It also exploits applications to provide machine-readable and unambiguous data service. In the future, the technologies from the semantic web and the cloud computing may be more tightly combined and leverage each other to provide intelligent and scalable services.

3.2 Ontology learning issue in web-scale semantic information processing

A building block of the semantic web construction is ontology techniques. Ontology comes from

Greek and used as a philosophical discipline as the science of existence or the study of being. When ontology is introduced in artificial intelligence to represent knowledge, Gruber has given the formal definition [17] : An ontology is a description (like a formal specification of a program) of the concepts and relationships that can formally exist for an agent or community of agents. An ontology can be used for vocabulary and taxonomy sharing and domain modeling with the definition concepts and their properties and relations. The W3C consortium recommended OWL for web-scale semantic information modelling, which is based on the Description Logic (DL) [18] formalism for knowledge representation and reasoning by ontology. Ontology techniques offer a direction towards solving the interoperability problems brought by semantic obstacles.

Though ontology techniques can help knowledge sharing, its learning and construction by human being is costly and erratic so that it makes ontology a challenging enterprise. To solve this problem, the study on semi-automatically extraction of the concepts and relations from data becomes hot. This process is referred as ontology learning [19]. From the type of input data, the ontology learning can be classified into two kinds: learning from structure or semi-structure data and learning from unstructured data [20]. The former kind of ontology learning is mainly mapping or lifting definition from the schema to corresponding ontological definitions, such from XML or database to OWL. The latter kind of ontology learning, it is also referred as ontology learning from text. A number of studies on ontology appeared to satisfy the ontology construction's needs. Text-to-Onto [21], a ontology learning tool, is able to learn concepts from unstructured or semi-structured data. It uses multi-strategy method to combine association rules, formal concept analysis and clustering. The web-kb [22], on the other hand, combines Bayesian learning and FOL rule learning to learn instance and rules from the Web. These ontology learning tools can significantly reduce the time and efforts in ontology construction compared to human work.

Though ontology learning can help build ontology model, the knowledge acquisition and modeling are often hard with the increasing information on the web. Since the information on the web is often in distributed and heterogeneous manner, there are always conflicts and ambiguities in ontology learning and reasoning from the web. Some researchers have build their ontology model from some shared vocabularies such as Wikipedia or folksonomies [23,24] to serve the knowledge sharing and acquire domain knowledge, but conflicts are still formidable problems when reasoning on these distributed ontologies. In summary, without sufficient strategy to solve the semantic conflicts and ambiguities in the distributed and heterogeneous web environment, the ontology model will not be able to help satisfactory knowledge acquiring and sharing.

3.3 Semantic service in the semantic web

In recent years, the semantic web technologies have been widely used in web service, data extraction and other domains to provide semantic unambiguous and knowledge sharing service. Cyc [25], for example, is a large foundation ontology for formal representation of the universe of discourse. In bio-science domain, gene ontology and disease ontologies are widely used to describe the biological and clinical investigations. Besides these single ontologies, ontology libraries have been developed and led to the services providing lists of directories of ontologies with search facility. The static libraries such as the protégé ontology library, and DAML ontology library contains a set of domain ontologies in various format. Some ontology libraries are also served as web search engines. Swoogle [26] and Falcon [27] can both search for RDF resources and ontologies available on the Web.

These semantic services have brought a large scale of data to be processed on the web. According to the statistics collected by the linked open data community, the Web of Data consists of 4.7 billion RDF triples, which are interlinked by around 142 million RDF links in 2009. How to effectively query and manage large scale of data becomes one of the most urgent challenges to semantic service in the semantic web. At the same time, with the emergence of the cloud computing services, the number of ontology and RDF data will obviously increase quickly, and the efficient and effective semantic service will become more and more important.

**4 Web-Scale semantic information processing for cloud computing**

There has been a tremendous increase in internet services in the past few years via the technologies of cloud computing. It is now well recognized that these internet service (e.g., virtual hardware renting, storage service, programming platforms and software service) contains valuable semantic information that can be exploited for many applications. Exploiting the semantic information more effectively via the use of the semantic model, parallel processing technique, and web service technology, more accurate and sophisticate cloud applications can be built. Sheth et al [28] have argued that the web-scale semantic information processing and modeling can be helpful in three aspects of cloud computing: functional and nonfunctional definitions, data modeling, as well as service descriptions, license, etc. Based on these enhanced applications, high quality and efficient cloud service can be generated. Many significant researches have been done to investigate new strategies available in cloud computing framework. In this section, we review some new strategies for semantic information processing in cloud computing.

4.1 Semantic information in cloud computing

The semantic information in cloud comprises various forms of metadata from web applications, software environments, communication devices, hardware, which contain both pure semantics and descriptive semantics. Therefore, it provides a huge potential to share deeper knowledge among users, service providers, hardware environment, software applications and encourage reuse and interoperations. It has become an important information resource in addition to program material. From the semantic information in the cloud, it is possible to acquire cloud service composition, work condition of the agents, or users' taste and requirements. The growing and readily available semantic information in cloud computing is rising the new opportunity to share knowledge widely and efficiently compared with the existing web techniques and to mitigate the large scale data processing and semantic conflict problems considerably.

The semantic information in cloud expresses the users' opinions and text meaning, as well as the applications' logic of the software, work flow, and hardware operations. The semantic information sharing and analysis such as semantic layers classification [29] and knowledge sharing in software engineering [30] are possibly as augmentations to cloud service, since it might help understanding between cloud service providers and users, or among different cloud service developers.

The service developers show their interest in semantic information about cloud resources discovery or intelligent service construction. They adopt semantic model to describe the domain knowledge and carry out distributed reasoning. The industry and academic are increasingly coming to realize that semantic information is crucial to construct satisfactory cloud services to the consumer and they are paying more and more attention to these issues. There are already many companies that focus on semantic information processing in cloud computing and examples include Yahoo and Google.

4.2 Ontology model in the cloud

As a formal conceptual semantic model for sharing knowledge, ontology can provide a unique semantic ground for services and applications. There are already many researches on the ontology model in the cloud. A unified ontology of cloud computing is proposed by Yourseff [31], demonstrates a dissection of the cloud into five main layers, and illustrates their inter-relations as well as their inter-dependency on preceding technologies. This unified ontology is envisioned as a stack of layers, and each layer encompasses one or more cloud services which have equivalent levels of abstraction. Besides the application and software, hardware and firmware are also included as the bottom layer of the cloud stack. The semantic information containing in these hardware services can facilitate the interoperation among different cloud systems and help the high layer in the cloud stack enhance their performance.

Jang et al. [32] designs a context model based on ontology in mobile cloud computing in order to provide distributed IT resources and services to users based on context-aware information. Generic ontologies are defined and connected through relational property, while domain ontologies make the concepts in generic ontology more detail. It can provide extensibility and formal representation ability by hierarchical ontology classification.

Haase et al. [33] studied the intelligent information management in enterprise clouds and proposed a eCloudManager ontology to describe concepts and relationships in enterprise cloud management. The eCloudManage ontology can easily be extended by the user to capture information relevant for specific use cases or to integrate other data sources. Zhou et al. [34] also design an enterprise software ontology to reengineering enterprise software for cloud computing. They used a case study of the Plazma business solution system to show that the ontology-based approach is an efficient reengineering methodology for large-scale software system in terms of automating.

These applications can help to represent the domain knowledge and provide a unique semantic ground to avoid the semantic ambiguity. However, since in cloud computing, resources are distributed and heterogeneous, the ontologies may be independently developed and are heterogeneous, too. These ontology heterogeneity, including synonyms, hyponyms, or structure heterogeneity can lead to semantic conflicts and unsatisfying in semantic reasoning. Another serious problem is that the traditional reasoning algorithms are not designed for distributed environment but for central processing [35]. All these disadvantages in cloud computing bring challenges to make use of ontologies to describe and share domain knowledge accurately and efficiently. Thus, how to solve these problems caused by the heterogeneity and distributions of the cloud computing is a key issue to improve the performance of the ontology model in cloud computing.

The work of [36] makes an alignment between ontologies in a cloud computing architecture. However, this work didn't consider reasoning among the distributed ontologies. As a distributed reasoning architecture, DRAGO [37] is based on local semantics [38] and uses distributed description logics framework [39] to represent multiple semantically connected ontologies. By constructing bridge rules, it provides reasoning service for multiple OWL ontologies interconnected via C-OWL [40] mapping and builds a sound and distributed tableau reasoning algorithm among C-OWL. However, reasoning based on distributed description logics will sacrifice expressiveness in the links between local models and drop some formal properties on the level of the overall model.

Different from DRAGO, the approach proposed in [41, 42] construct a distributed, complete and terminating algorithm that decides satisfiability of terminologies in $\mathcal{ALC}$ and guarantees the global semantics to be preserved. The algorithm extends ordered resolution with a method for assigning clause to a unique location at which all possible resolution steps are executed by a local solver. Because this resolution for $\mathcal{ALC}$ can be distributed across different distributed reasoners, this method can reason upon interlinked ontologies in cloud computing.

4.3 Parallel processing

The web-scale semantic information are in large corpuses and grows fast so that it is hardly to be processed and stored by a single processor. Especially, the semantic rules are often represented in different forms that require preprocessing and coordination among reasoners [43] in the cloud computing environment. The parallel framework in cloud computing can be adopted to solve this problem and there has been a lot of work that deals with web-scale semantic information processing the cloud. The parallel processing on the web-scale semantic information mainly contains parallel storage and query, as well as parallel reasoning.

*4.3.1 Parallel storage and querying*

Though the state-of-the-art distributed database such as HDFS [44] can store and manage the web-scale semantic information like to other data, there are still many works on how to make these storage and query more speed-up and scalable. For parallel storaging and querying web-scale of RDF triples, Yahoo builds grid computing using hadoop in a batch-processing model using clusters of hundred of machines. It applies the parallel database Pig [11] to support for data transformations such as projections grouping, sorting, joining, and compositions. However, Pig provides no index and requires full parsing when updating.

Different from pig, YARS2 [45] uses six different indexes into six data orderings, supporting the Semantic Web Search Engine (SWSE) [46] to query and process the large scale semantic information efficiently in parallel fashion.. It makes a binary search upon the index and retrieves the closest block of data. In order to distribute its indexes, YARS2 uses a hash partitioning over the first attribute of the quad. This mechanism may cause disadvantages when considering data orderings that are predicate-first. Hence YARS2 randomly distributes predicate-first orderings and floods queries that require this ordering to all machines. Similar to YARS2, Virtuoso [47] also makes hash partitioning to split its data and indexes. But Virtuoso is based more on a object-relational DBMS heavily optimized for RDF storage than the heavily read optimized federated repository YARS2 does. Virtuoso performs rebalancing by moving responsibility and relevant data for certain virtual machines from one physical machine to another. Rebalancing is performed on-the-fly but time-consuming.

The clustered TDB [48] builds the clustered RDF triple store based on Jena TDB, which is composed of query coordinator and data node. Query coordinator can receive queries and producing query plans, as well as control execution on the data nodes. The data node is responsible for data storing and perform operations such as sorts, join, and so on. There are three forms of parallelism in querying performed by the clustered TDB: inter-query, intra-query and intra-operation. The authors focus their parallelizing operations mainly on pipelining and partitioning. The evaluation results show that clustered TDB offers near linear scaling characteristics in load times and can be speedup and scaleup.

*4.3.2 Parallel reasoning*

A vast literature on scalable and parallel reasoning has grown in these areas of research. Some preliminary work has shown the benefits of utilizing subset of the first order logic semantics in parallization making [49].

DORS [50] combines description logic reasoned for TBox reasoning and rule engines for Abox reasoning based on distributed hashtable and relational database. When data is added, the TBox reasoned materializes all TBox triples, which are then reasoned by ABox reasoned. Kauodi et al [51] also build the forward-chaining and backward-chaining to perform RDF inferencing and querying on top of DHTs. They stored each triple in three places based on hashing of subject, predicate and object. However, there are load-balancing problems since term popularity in RDF exhibits a power-law distribution [52].

Newman et al [53] decompose and merge RDF graphs in order to distributed and query RDF triples on top of MapReduce and hadoop. They extended the definition of RDF molecules to include hierarchy and ordering to disambiguate blank nodes. They used BioMANTA ontology [54] as their testbed and perform SPARQL queries on the data. They claimed that their method can help to alleviate the co-identification [55] problems and enhance scalability of RDF storage and query.

In paper [56], OWL horst semantics [57] are used in generating the rule set for reasoning. Data partitioning and rule partitioning are used to partitioning the workload and hence parallelize it. In the former partitioning, every process gets a fraction of the data and all the rules, while in the latter, every process get all the data and a fraction of the rules. The results show that the rule-partitioning is cheaper than the data partitioning and has better speedup when the data-sets are dense graphs.

SAOR [58] implemented a fragment of OWL horst semantics to allow efficient materialization and prevent "ontology hijacking". It uses a single machine to compute the closure of an RDF graph and later extend the approach to support a subset of OWL 2RL for distributed reasoning on 1.1 billion triples.

Willioms et al [59] presented a straight forward parallel RDFs reasoning on a cluster. They used Abox partitioning safe rules which ignore the RDFs schema extending to ensure the independency of the partitioning. Each processor calculates its single chunk and the results are merged. The approach does not support complete RDF reasoning. Also, no global data structure is allowed to index the input data to reduce inter-process communication cost.

In paper Liebig [60], a nondeterministic tableaux is exploited. This work concerned with disjunction rule and number restriction merge rule since there are no dependencies between the alternatives and hence these rules can be evaluated within parallel threads. The reasoning is performed on ABox and the results show a limited scalability.

The large knowledge collider (larkc) [61] is an open architecture and a generic platform for massive distributed reasoning tasks and allows components (plug-ins) responsible for diverse processing tasks in each work flow. As a part of larkc, Marvin [62] is a parallel and distributed platform for processing large amounts of RDF data. The authors presented a divide-conquer-swap based on data-partitioning in a p2p network. A load-balanced auto-partitioning approach was used without upfront partitioning costs. The experimental results show Marvin can calculate the closure of 200M triples on 64 compute nodes in 7.2 minutes.

WebPie (web-scale parallel inference engine) [63, 64] performs parallel rule-based forward reasoning based on Mapreduce framework. WebPie introduced several optimizations to improve the performance on RDFS and OWL-horst reasoning. Their experimental results on LDSR and LUBM show that WebPie can process large-cale of RDF triples in very high speed. It computes the closure of the 1 billion triples in less than 1 hour using 22 machines. WebPie shows the penitential high efficiency of parallel reasoning on web-scale semantic information by using cloud computing techniques.

Inspired by webpie, Mutharaju et al [65] present a parallel algorithm for classifying $\mathcal{EL}^+$ ontologies using Mapreduce. They used a set of completion rule and the CEL algorithm to perform classification of $\mathcal{EL}^+$ ontologies. The TBox is distributed on different node of the cluster. The paper shows that the Mapreduce frame work can help parallel classification of $\mathcal{EL}^+$ ontologies soundly and completely.

Mina & Haarslev [66] also propose a new algorithm for TBox classification. A multi-threaded architecture providing control parameters and the partree is constructed for each thread. This work is still on-going and only limited experimental results are reported.

Bao et al [67] describe a distributed reasoning algorithm for P-DL and adopts a federated approach to reasoning with modular ontologies by local reasoned. The local reasoners communicate with each other in a asynchronous fashion.

Though parallel reasoning in the cloud computing is still a open question, from these studies we can assume that the cloud computing techniques can at least partially solve this problem and help to improve the efficiency and scalability.

4.4 Semantic service in cloud computing

Before the cloud computing service has been mentioned, the semantic web service has been widely adopted to help providing intelligent machine-to-machine interaction over a network. The semantic web service languages include the Web Services Semantic (WSDL-S), SAWSDL, OWL, while the frameworks are mainly composed of OWL-S, WSMF, WSMO, etc. These semantic web service techniques can be combined with the cloud computing framework to get scalable and unambiguous semantic cloud service. There are also a lot of work deal with ontology model application and semantic web services adoption in the cloud.

SITIO [68] deals with the integration of heterogeneous applications by adding machine-understandable and machine-processable metadata to web resources through ontologies. It used a semi-automatically semantic annotation by mapping domain ontologies and the web services. SITIO can be adopted as a business process based on semantic platform where services are executed from a SaaS perspective. Kim et al [69] presents an e-portfolio system design based on private-public data index system PrPl. PrPl is constructed on a personal cloud architecture and maintains a semantic index which allows semantic query based on attribution. Users can access the semantic services through a visual map-based e-portfolio interface.

Besides these domain ontologies and semantic index, there has been a lot of work that deals with cloud service discovery. Cloudle [70] compares the relationships among cloud services through concept similarity, object property similarity and data type property similarity. The authors also build a agent-based discovery system that consults an ontology when retrieving information about cloud service. The results show that the ontology-based cloud service search is more precise and intelligent. Dastjerdi et al [71] also proposed a flexible approach for performing ontology-based discovery of cloud virtual units. The virtual units in the IaaS are modeled into web service modeling ontology (WSMO). Users can find the best suited cloud service provider using the ontology discovery.

Some researchers have found that semantic middleware can leverage the performs of the cloud services. Liu et al [72] describes a semantically-enhanced scientific resource library portlet to enable interaction with multiple distributed repositories in the cloud based on a semantic context middleware Tupelo. The semantic resource library is composed of three layers: web 2.0 interaction layer, semantic context union and mapping layer, and data store layer. The model is based on RDF and distributed over the virtual machine in the cloud. Tupelo middleware is used to implement on RDF and content abstraction over a wide range of triple stores, databases, and file systems. By leveraging rich metadata available to different data type, users are able to explore all kinds of digital resources and they relationships by a broader graph model.

Yang et al [73] also employed a backend information agent system ontoIAS to design a ubiquitous interface agent in the cloud. ontoIAS stores an ontological database to provide ontology-directed canonical format for storing webpage information to avoid numerous, jumbled in correct information torrents. Users can employ the ubiquitous interface agent to use ontoIAS via related mobile equipment.

Apart from these ontology-based semantic cloud services, the web-scale semantic information processing has been adopted directly in the non-central and distributed cloud computing environment to improve the performance. For example, Delta-Reasoner [74] is semantic reasoner built on mobile platform to provide intelligent mobile service. By combining information obtained via device sensors with background knowledge, Delta-Reasoner can deduce the users current context to adapt the application's behaviour to the user's needs. As a document-oriented lookup index for open linked data, Sindice [75] adopts Hadoop to build a parallel architecture to provide indexing and querying services in the cloud computing environment. In light of these studies, it can be said that the semantic information and formats can be employed to supplement current cloud services.

**5 Conclusion**

The purpose of this survey has been to describe and analyze the state-of-the-art of the web-scale semantic information processing in cloud computing environment, and to convey to the reader a sense of our excitement about the intellectual richness and breadth of the area.

The web-scale semantic information processing contains the key to solve the semantic ambiguity that hinders the leverage of precise cloud service and knowledge discovery. The semantic web is an emerging research direction fulfilling various semantic processing tasks. Many semantic web techniques and ontology model have been used to underpin an semantic cloud service. The applications of cloud ontology, parallel semantic processing and semantic cloud service have been covered in this study.

In the recent gears, many research groups have invested much effort on semantic cloud service, and have made many great achievements. The rich literature is growing around these topics. However, challenging problems still exist in these areas. In particular, the research issue on how to make parallel inference, how to adopted ontologies in the distributed cloud computing environment have attracted lots of the research attentions. We very much hope we have provided some helpful information to the readers who are encouraged to take up the many challenges that remain in the area.

## 6 Acknowledgement

Thanks a lot for the help of the anonymous reviewers.